\documentclass{ws-ijmpc}

\usepackage{graphicx}
\usepackage{dcolumn}
\usepackage{bm}

\usepackage{amsmath}
\usepackage{amssymb}
\usepackage{url}
\usepackage{subfigure}
\usepackage{algorithm}
\usepackage{algorithmic}
\usepackage{enumerate}
\usepackage{slashbox}
\usepackage{mathrsfs}
\usepackage{CJK}
\usepackage{algorithm}
\usepackage{algorithmic}

\begin{document}

\markboth{H.M.Cheng, etc.}
{Community Detection in Complex Networks using Link Prediction}

%
%

\title{Community Detection in Complex Networks using Link Prediction}

\author{Hui-Min Cheng, Yi-Zi Ning, Zhao Yin, Chao Yan, Xin Liu, Zhong-Yuan Zhang*}

\address{School of Statistics and Mathematics, Central University of Finance and Economics,\\ Beijing, China\\
zhyuanzh@gmail.com }

\maketitle

\begin{history}
\received{Day Month Year}
\revised{Day Month Year}
\end{history}

\begin{abstract}
Community detection and link prediction are both of great significance in network analysis, which provide
very valuable insights into topological structures of the network from different perspectives. In this paper, we propose a novel community detection algorithm with inclusion of link prediction, motivated by the question whether link prediction can be devoted to improving the accuracy of community partition. For link prediction,  we propose two novel indices to compute the similarity between each pair of nodes, one of which aims to add missing links, and the other tries to remove spurious edges.
Extensive experiments are conducted on  benchmark data sets, and the results of our proposed algorithm  are compared
with two classes of baseline.
In conclusion, our proposed algorithm is competitive, revealing that link prediction does improve the precision of community detection.
\end{abstract}


\section{Introduction}\label{Introduction}
Research on complex networks has become increasingly popular in many scientific disciplines, including sociology, transportation, and biology, etc\cite{strogatz2001exploring,wasserman1994social,girvan2002community}. Past works of complex network involve two principal lines of research\cite{chen2012introduction},  community detection and link prediction, both catching increasing attention due to their theoretical significance and potential application\cite{lu2011link,clauset2008hierarchical,fortunato2010community}.

In most cases, community is  characterized as a set of nodes which are densely connected internally and have comparatively sparser connections with the rest \cite{fortunato2010community,zhang2013overlapping}.
 Detecting community is of great significance to unveil the topological structures   of networks\cite{you2016community,zhang2015community}, for instance, communities in a social network can correspond to groups with similar interest or same goal\cite{moody2003structural,krishnamurthy2000network}, while in biology they might represent tissues with related function\cite{girvan2002community,sharan2006modeling}.
Community detection has many concrete applications, such as mining customers with same purchasing interests with the aim to provide better services or to recommend precise commodity on the internet\cite{reddy2002graph}.

However, most of real-world network data are incomplete and even inaccurate, resulting in missing  and even spurious links.
For many networks, such as metabolic networks and social networks, it is very costly and even impossible to check all potential interactions.
Instead, link prediction seeks to identify missing interactions, spurious edges, and predict future links by estimating the possibility of interactions between two nodes. Thus, link prediction plays a pivotal role in analyzing networks with missing  and spurious links, and is of considerable value in sense of application, such as recommending promising friends and reducing experimental costs.

The studies of link prediction and community detection are supposed to be mutually beneficial.
With community structures taken into account, the precision of similarity-based link prediction algorithm is increased\cite{soundarajan2012using,valverde2013exploiting}, illustrating that in-depth understanding of community information can be devoted to improving the accuracy of link prediction.
However, to the best of our knowledge, there is no work analyzing whether the accuracy of community structures detection can also be improved by link prediction methods. Traditional community detection algorithms  focus only on the raw network  structure, however, most of the available network information is incomplete, and  link prediction can be used to  approach the true network structure.

Here, we put forward a question: whether link prediction can be devoted to improving the accuracy of community detection?
In other words, the question we pose is that whether the results of community detection algorithm make difference between raw network  and predicted network.
Intuitively, more similar the given network is to true network structure, more akin its detected community is to ground-truth community structures.
Naturally we are motivated to pose a novel method, \underline{C}ommunity
\underline{D}etection using \underline{L}ink \underline{P}rediction (CDLP for short), which not only helps to detect  community information more accurately, but also throws light on the relationship between community detection and link prediction.

The rest of this paper is organized as follows: Sect. \ref{rw} sheds light on previous related work. Sect. \ref{description} illustrates the proposed method CDLP in details. Sect. \ref{experiment} gives the experimental results. Finally, Sect. \ref{conclusion} concludes.

\section{\label{rw}Related Work}
In this section, we introduce some relevant works briefly, including  link prediction and community detection methods.
\subsection{Link prediction}
 Previous algorithms of link prediction can be categorized into three classes\cite{lu2011link}: algorithms based on similarity score, algorithms based on maximum likelihood estimation and probabilistic models.  The algorithms based on maximum likelihood estimation and probabilistic models both try to best fit the observed data by estimating a group of parameters, former of which considers structural characteristics,
 such as hierarchical organization and community structures\cite{clauset2008hierarchical}, while the latter
 attemps to model joint distribution by optimizing target function, and make prediction by estimating the conditional probability\cite{clauset2008hierarchical}.
 However, the result is disappointing when they deal with big network
 with tens of millions nodes\cite{friedman1999learning}.

Similarity-based algorithms are the most commonly used methods, which is our main focus. Given a network $\mathscr{G}$, different similarity-based algorithms assign pairs of nodes different scores based on observed information,
including attributes of nodes and network structure.
In this case, higher score of yet nonexistent link indicates higher likelihood  to be added, while lower score of observed edge suggests higher possibility to be fake. The main problem  here is  how to quantify the value of nodes similarity. According to the information used, algorithms can be further classified into three categories: local indices, global indices and quasi-local indices. Compared with algorithms using global topology information, those based on local information generally have less time complexity with sacrifice of accuracy\cite{liben2007link}. In addition, there are many local similarity indices. We are concerned with the question "can we find some good local indices that meet the requirements of quality and speed?". Emperical experimental results showed that the simplest neighborhood-based index, namely Common Neighbours index (CN for short) has good performance\cite{zhou2009predicting}. For this reason, we choose CN metric as "base metric" for our proposed similarity metrics.

For one pair of nodes (a,b) in a given network, the value of CN is computed as:
\begin{equation}\label{}
S(a,b)=\left|\Gamma(a, b)\right|,
\end{equation}
where $\Gamma(a, b)$ is the set of common neighbours of nodes $a$ and $b$. Intuitively, more common neighbours two nodes have, more likely they are to have a link\cite{lu2011link}. The metric has been widely applied to study social networks, where individuals who have many mutual friends are suggested to be friends in the future\cite{kossinets2006effects}.

\subsection{Community detection}
 Community detection  aims to divide networks into modules, meaning groups of densely connected nodes\cite{girvan2002community,newman2006modularity}.  Over the years, the problem of community detection has been studied by many researchers, and many methods have been developed, such as graph partitioning, spectral clustering and hierarchical clustering, etc\cite{newman2004detecting,go2015statistical}.
 To evaluate how good an algorithm is, Newman and Girvan proposed modularity\cite{newman2004fast}, which is a prominent and most commonly used measurement determining the quality of community partition. For a given network  $\mathscr{G}$, modularity measures the internal connectivity of identified communities with reference to a null model, which is a randomized graph with exactly the same node-degree sequence regardless of community structures. Suppose a given network $\mathscr{G}$ containing $N$ vertices and $M$ edges, the modularity function $Q$ is defined as\cite{chen2012introduction}:
\begin{equation}\label{modularity}
Q=\frac{\sum \limits_{i,j} \left(a_{ij}-p_{ij}\right)\delta\left(C_{i},C_{j}\right)}{2M},
\end{equation}
where $M$ represents the total number of edges in the network $\mathscr{G}$ , $a_{ij}$ and $p_{ij}$ denote the real and expected number of edges between node $i$ and $j$ respectively, while $C_{i}$ and $C_{j}$ indicate the communities to which nodes $i$ and $j$ belong, respectively, and $\delta\left(C_{i},C_{j}\right)$ is set to 1 if the nodes $i$ and $j$ are in the same community and 0 otherwise.
The problem of how to calculate the expected number of edges connecting node $i$ and $j$ is tackled by the definition as follows:
\begin{equation}\label{expected}
p_{ij}=\frac{k_{i}k_{j}}{M}.
\end{equation}
Thus, $Q$ is equal to 0 if the whole network is considered as one community, and higher value of $Q$ indicates stronger community structures. In extreme cases, an ideal partition structure, where communities are independent with each other, yields a modularity value of 1.

Since higher value of  $Q$ corresponds to higher quality of community detection, can we find the best partition by simply optimizing $Q$?
Based on the idea of this,  fast greedy modularity optimization algorithm is proposed to  find the optimal community partition by directly implementing a greedy optimization method in agglomerative approach\cite{newman2004fast}\cite{lancichinetti2009community}.

The main idea of fast-greedy algorithm is as follows: In initial step, it assigns each node to one community, and these communities are agglomerated step by step. The algorithm is stopped when all nodes are combined into one community. At each step, greedy principle is applied as merging criterion. In other words,
at each step, community structures whose amalgamation gives the largest increase or smallest decrease in modularity will be chosen\cite{newman2004detecting}
\cite{orman2011accuracy}.
By comparing  $\{Q_{1},Q_{2},...,Q_{n} \}$, of which $Q_{m}$  corresponds to the modularity value at the step of $m$, we will select the best one  at the state with $Q= \max \{Q_{1},Q_{2},...,Q_{n} \}$. Fast-greedy algorithm is widely used for community detection, and is adopted as baseline method in this paper.

\section{\label{description}CDLP Method Description}
In this part, we begin with description of  components in the method, and then  we describe the CDLP algorithm in details.

\subsection{Proposed similarity measurement}

In this paper, we propose two new indices to measure similarities among pairs of nodes, in order to add new edges and delete existent edges separately.

\subsubsection{$A$ index}
Calculation of $A$ index depends on the  community membership information.
With community structures obtained, $A$ index estimates the probability of a new edge creation between two nodes in same community, but not yet linked.
Consider each pair of non-linked nodes, $a$ and $b$, which are in the same community $k$,
$A$ index is defined as follows and assigns high score to those whose most of common neighbours are in the same community:

\begin{equation}\label{A}
A(a,b)=\frac{2\sum \limits_{i\in\Gamma(a,b)}\left|C(a)\cap C(b)\cap C(i)\right|}{d(a)+d(b)},
\end{equation}

where $\Gamma(a,b)$ is the set of common neighbours of nodes $a$ and $b$, $C(a)$ is the community which node $a$ belong to. And $\left|C(a)\cap C(b)\cap C(i)\right|$ is $1$ if nodes $a$, $b$ and $i$ are in the same community, and 0 otherwise. $d(a)$ is the degree of node $a$.

\begin{figure*}
\hspace*{-10mm}
\includegraphics[height=90mm,width=170mm]{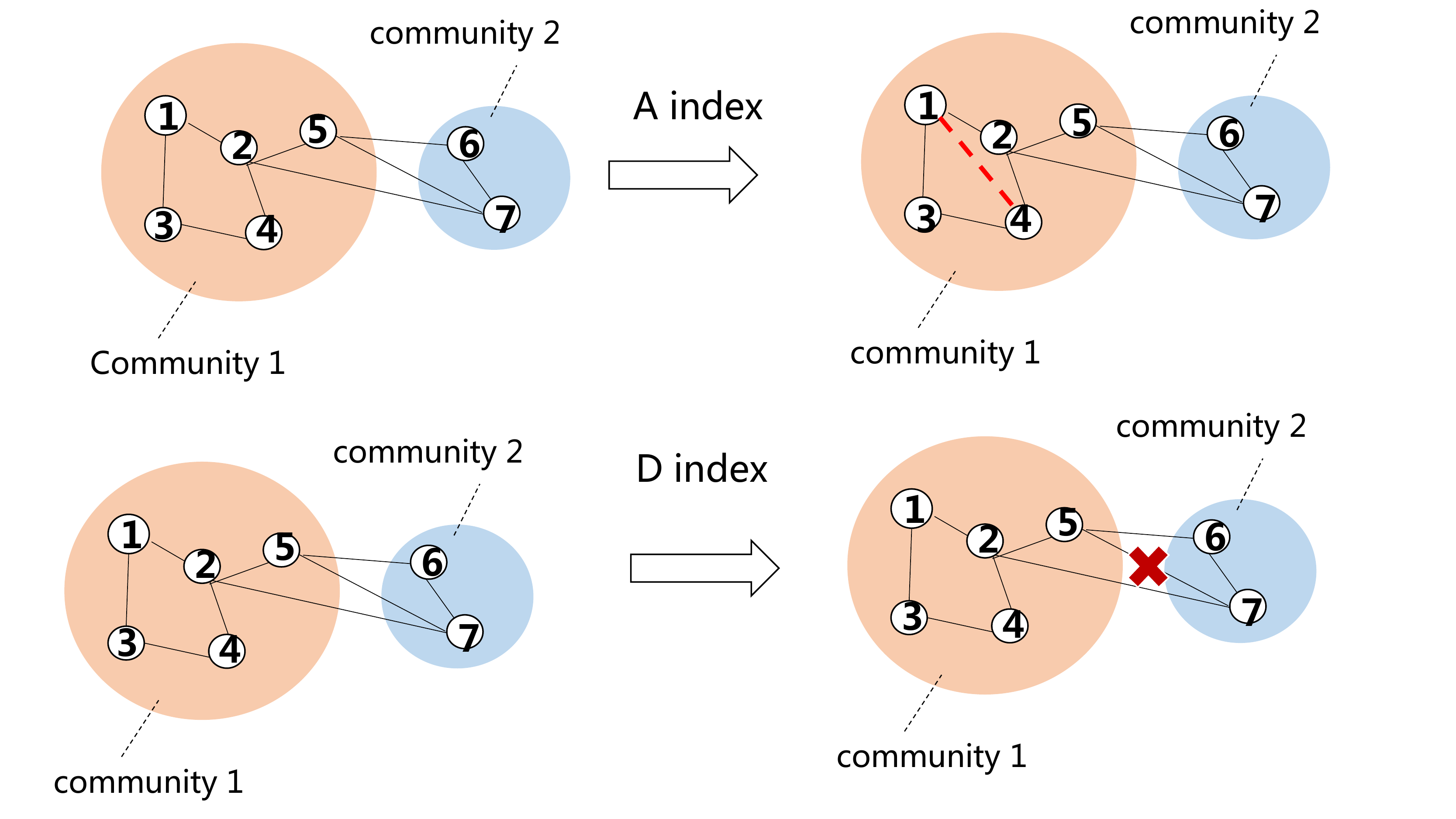}
\caption{An exemplification illustrating the calculation of $A$ index and $D$ index}\label{Fig:example1}
\end{figure*}

To explain this index more clearly, we take network with two communities  in Fig. \ref{Fig:example1} as an example. $I$ is the set of all pairs of nodes which appear in the same community but do not have linkage yet. $S$ is the set of $A$ scores for $I$. For this network, $I = \{(1,4),(1,5),(2,3),(3,5),(4,5)\}$, and $S=\{1,0.4,0.67,0,0.4\}$. Clearly, higher score means higher likelihood of existence. With these scores ranked by decreasing order, we take top-$L$ edges as newly created links. Specifically, if we set $L$ as 1, then the predicted link is $(1,4)$, which is shown as red dotted line in Fig. \ref{Fig:example1}.

\subsubsection{$D$ index}
Analogously, $D$ index also depends on community membership information, but $D$ index is proposed to recognize spurious links.
With community structures obtained, $D$ index estimates the the probability of removing an existent edge between pair nodes in different communities.
Consider each linked pair of nodes, $a$ and $b$, which belong to different communities $k$  and $g$ respectively,
$D$ is defined as follows and assigns low similarity score to those whose most of common neighbours are neither in $k$ nor in $g$:

\begin{equation}\label{D}
D(a,b)=\frac{max\{\sum \limits_{i\in\Gamma(a,b)}\left|C(a) \cap C(i)\right|,
\sum \limits_{i\in\Gamma(a,b)}\left|C(b)\cap C(i)\right|\}}
{\left| \Gamma(a,b) \right|},
\end{equation}

where $\Gamma(a)$ is the set of neighbours of node $a$. $\left|C(a)\cap C(i)\right|$ is $1$ if nodes $a$ and $i$are in the same community, and 0 otherwise.

 We take the network with two communities in Fig. \ref{Fig:example1} as an example. $I$ is the set of all pairs of nodes which appear in different communities but have linkages. $S$ is the set of $D$ scores for $I$. For this network, $I=\{(5,7),(5,6),(2,7)\}$, and $S=\{2.5,5,5\}$. Lower score means higher probability of spurious edges. With these scores ranked by increasing order, we select top-$L$ ones. Specifically, if we set the $L$ as 1, then the removed link is $(5,7)$, which is presented by red cross in Fig. \ref{Fig:example1}.

\subsection{Algorithm description}
Given a network $\mathscr{G}$, we take three succeeding steps of link prediction. In each link prediction step, we apply $D$, $A$ and $D$ index respectively, to get predicted network $\mathscr{G}_1$, $\mathscr{G}_2$ and $\mathscr{G}_3$. For each predicted network, we also evaluate its community structures detected by fast-greedy algorithm using modularity $Q$. And the optimal predicted network with the highest modularity value is selected.
Detailed algorithm procedure is illustrated in Algorithm \ref{alg:Framwork}.

\begin{algorithm}[htb]
\caption{ CDLP}
\label{alg:Framwork}
\begin{algorithmic}[1] 
\REQUIRE ~~\\ 
Adjacency matrix $A$ of input network $\mathscr{G}$;\\
Proportion of nonexistent edges to be added, $P_{D}$;\\
Proportion of existent edges to be removed, $P_{A}$;\\
\ENSURE ~~\\ 
Community information $M$;

\STATE Change the network topological structure of raw network  $\mathscr{G}$ by removing some existent edges  determined by parameter $p_{D}$, using simmilarity matrix computed by $D$ index, to get predicted network $\mathscr{G}_1$;
\label{code:fram:trainbase}
\STATE Change the network topological structure of network  $\mathscr{G}_1$ by adding some nonexistent edges determined by parameter $p_{A}$, using similarity matrix computed by $A$ index, to get predicted network $\mathscr{G}_2$;
\label{code:fram:add}
\STATE Change the network topological structure of network  $\mathscr{G}_2$ by removing some existent edges determined by parameter $p_{D}$, using similarity matrix computed by $D$ index,  to get predicted network $\mathscr{G}_3$;
\label{code:fram:classify}
\STATE Compute modularity value in each step form 1 to 3, choose the optimal community structures  ${M}$ with the highest value.
\label{code:fram:select}
\RETURN ${M}$; 
\end{algorithmic}
\end{algorithm}

\begin{figure*}
\hspace*{-10mm}
\includegraphics[height=100mm,width=170mm]{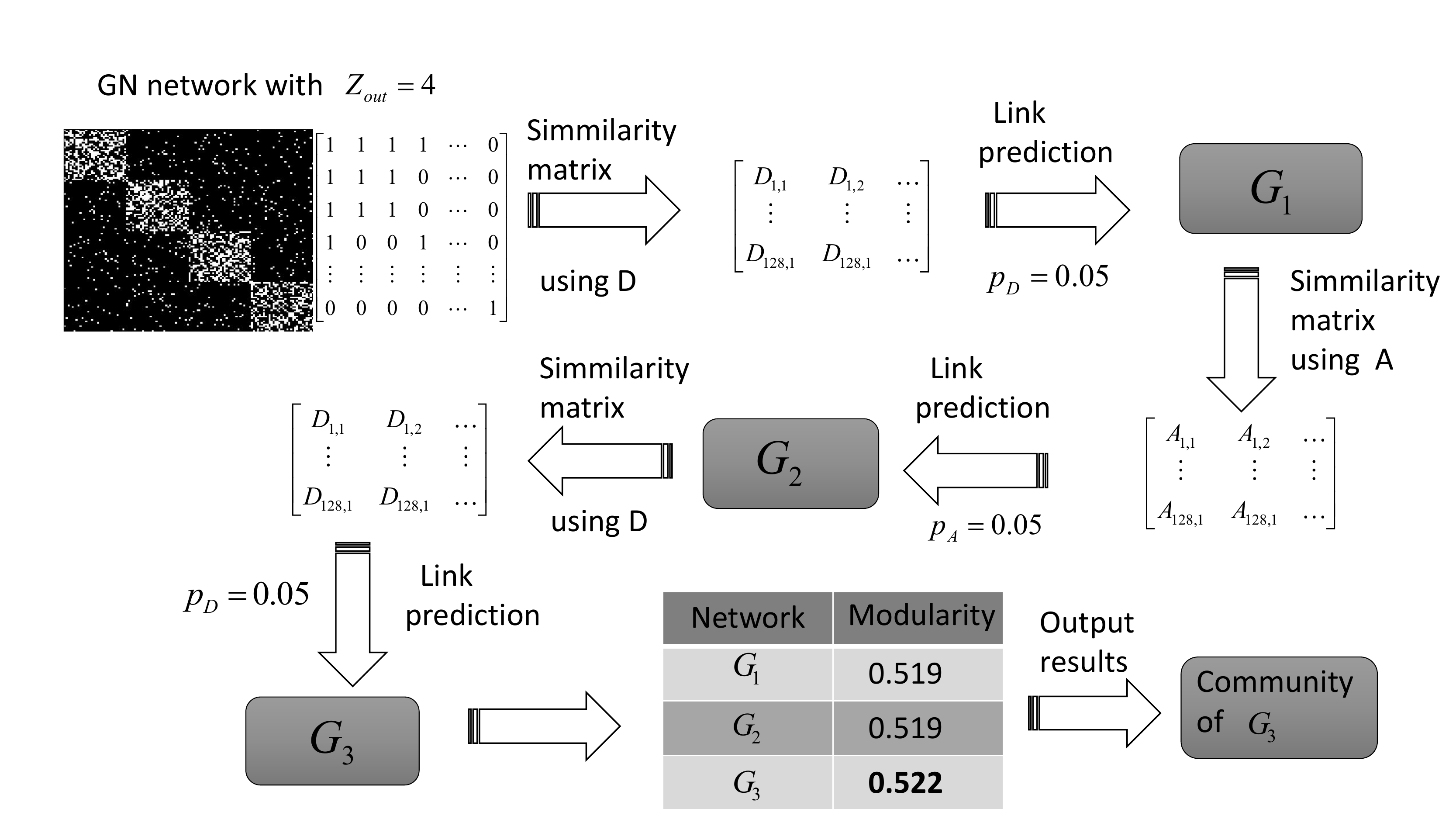}
\caption{Flowchart to illustrate how our proposed algorithm works, with GN network with $Z_{out}$=4, $P_{D}$=0.05 and $P_{A}$=0.05 taken for example.
}\label{Fig:flowchart}
\end{figure*}

To present our method more clearly, we plot specific algorithm flowchart as shown in Fig. \ref{Fig:flowchart}.
Firstly, we build three  similarity matrix to get three predicted network succeedingly, $\mathscr{G}_1$, $\mathscr{G}_2$ and $\mathscr{G}_3$.
For each predicted network, we compute its modularity value, using fast-greedy algorithm.
Finally, the corresponding community structures     with the optimal modularity value are output.

\section{\label{experiment}Experimental Results}
\subsection{Datasets}
To test the effectiveness of our proposed algorithm, we work with two classes of artificial networks, GN networks and LFR networks, both with precisely known community structures.
\subsubsection{GN network}
Girvan and Newman introduced a series of computer generated network GN for community detection\cite{girvan2002community}. For GN networks, all nodes are partitioned into 4 groups with equal size of 32 nodes. The average degree is 16, and the expected internal degree and external degree are $Z_{in}$ and $Z_{out}$, respectively. Obviously, as $Z_{out}$ increases, the detection of community structures becomes more challenging. In this paper, we generate GN networks with $Z_{out}$ ranging from 1 to 12.
\subsubsection{LFR network}
However, GN are small networks with homogeneous community size and node degree, which is not consistent with features of real world ones. To solve this problem, the Lancichinetti-Fortunato-Radicchi (LFR) benchmark networks with heterogeneous community size are proposed\cite{lancichinetti2008benchmark}. The networks are generated by several parameters, including number of total nodes denoted by $N$, average degree and max degree denoted by $k$ and $k_{max}$ respectively, exponent of power-law distributions of nodes degree and community size denoted by  $\gamma$ and  $\beta$, respectively, and a mixing ratio of external links denoted by $\mu$, which has the strongest effect on the algorithms performance. Specifically, as $\mu$ increases, the community structures become more ambiguous and more difficult to be detected. In this paper, we generate LFR networks with variable $\mu$ ranging from 0.1 to 0.9 and the constant parameters are: $N$=1000, $k$ = 20, $k_{max}$ = 50, $\gamma$  = 2, $\beta$ = 1.

\subsection{Metric for evaluation}

 In this paper, we adopt NMI metric to evaluate the effectiveness of our proposed algorithm. Given an artificial network with $n$ nodes and $k$ communities, the normalized mutual information (NMI) is defined as follows\cite{strehl2002cluster}:
\begin{equation}\label{expected}
I(M_{1},M_{2})=\frac{\displaystyle\sum_{i=1}^k\sum_{j=1}^k n_{ij}log\frac{n_{ij}n}{n_{i}^{(1)}n_{j}^{(2)}}}{\sqrt{\left(
\sum_{i=1}^k n_{i}^{(1)}log\frac{n_{i}^{(1)}}{n}\right)
\left(\sum_{j=1}^k n_{j}^{(2)}log\frac{n_{j}^{(2)}}{n}
\right)
}},
\end{equation}
where $M_{1}$ and $M_{2}$  can be interpreted as ground-truth and computed community partition respectively, $n_{i}^{(1)}$ and $n_{j}^{(2)}$ indicate the community size of actual community $i$ and computed community $j$ respectively, while $n_{ij}$ calculates the number of nodes falling into computed community $j$, which actually belong to ground-truth community $i$. The value of NMI ranges from 0 to 1, and the larger value indicates better partition results. By definition, a perfect partition structure, where every node is assigned to the right community, yields a value of 1.

\subsection{Baseline for comparison}
 In this paper, we consider  two classes of baseline methods for comparison, including:
\begin{enumerate}
\item $Baseline\_{1}$: Community detection methods without network structure changed. Here we apply fast-greedy algorithm as the first class of baseline;

\item $Baseline\_{2}$: Community detection methods with inclusion of link prediction. Here we adopt CN index in the second class of baseline.
\end{enumerate}
 The former type of baseline considers only the raw network,  While the latter allows for recognization of  wrong edges using CN similarity measurement.
 Firstly,  we compare results of $Baseline\_{1}$ with $Baseline\_{2}$. For comparison, we focus on whether link prediction can be devoted to improving accuracy of community detection. Secondly, we compare results of $Baseline\_{2}$ with our proposed method $CDLP$.
 For comparison, we concentrate on whether our proposed indices used for link prediction outperform  CN.

\subsection{\label{sensitivity}Sensitivity to parameters $p_{D}$ and $p_{A}$}
\begin{figure*}
\hspace*{-10mm}
\includegraphics[height=60mm,width=170mm]{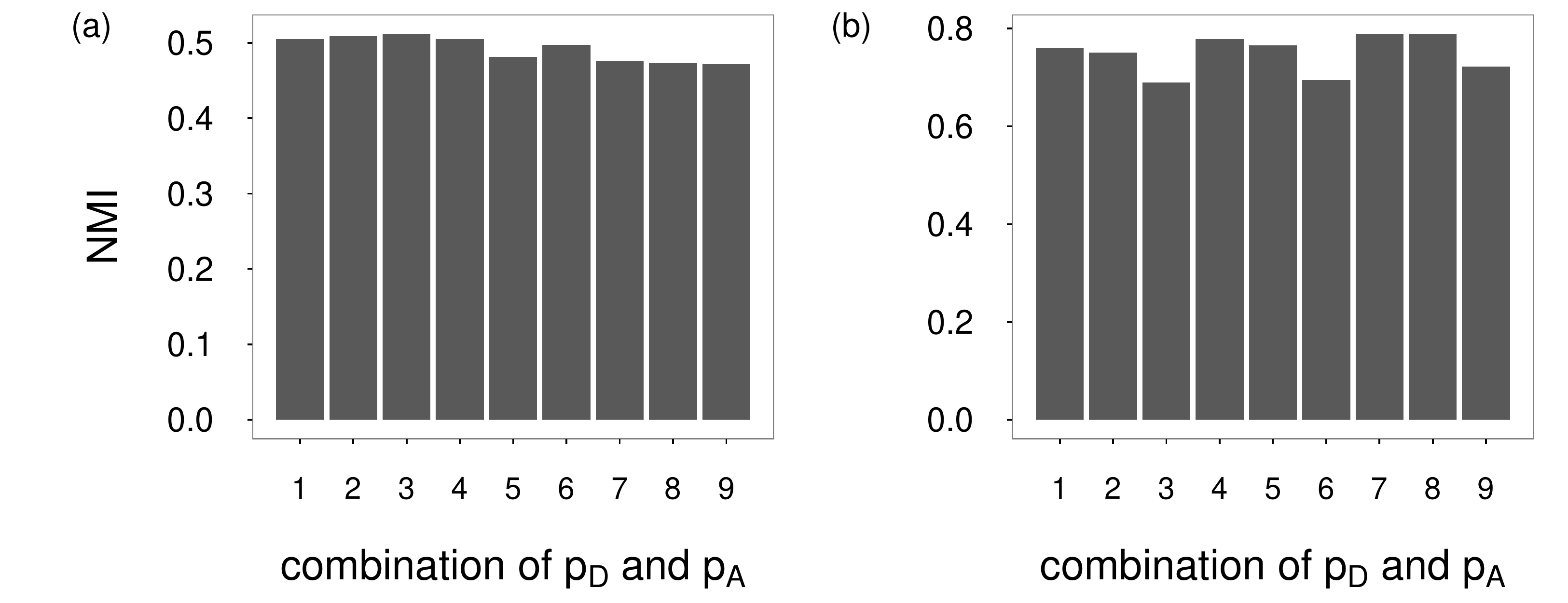}
\caption{NMI value of our proposed algorithm with different input combinations of  $p_{D}$ and $p_{A}$: (a) GN network with $Z_{out}$= 8, and (b) LFR network with μ = 0.5. There are 9 combinations of $p_{D}$ and $p_{A}$, since both of them  are chosen from three numbers, that is, 0.05, 0.1 and 0.15. }\label{Fig:sensitivity}
\end{figure*}

In order to investigate whether variations in the input parameters will cause a significant fluctuation of the output results, we test our proposed algorithm
with different input combinations of  $p_{D}$ and $p_{A}$  on the GN and LFR networks. We take two representative networks for comparison, GN with $Z_{out}$ = 8, and LFR with $\mu$ = 0.5. As is shown in Fig. \ref{Fig:sensitivity}, we can conclude that in both GN and LFR networks, our proposed algorithm is insensitive
to the parameters $p_{D}$ and $p_{A}$.

\subsection{\label{results}Results and analysis}
In this section, we demonstrate the results of numeric experiments. Since we have true community structures in mind, we apply the NMI measure to  evaluate the accuracy of our proposed algorithm.

\begin{figure*}
\hspace*{-10mm}
\includegraphics[height=100mm,width=170mm]{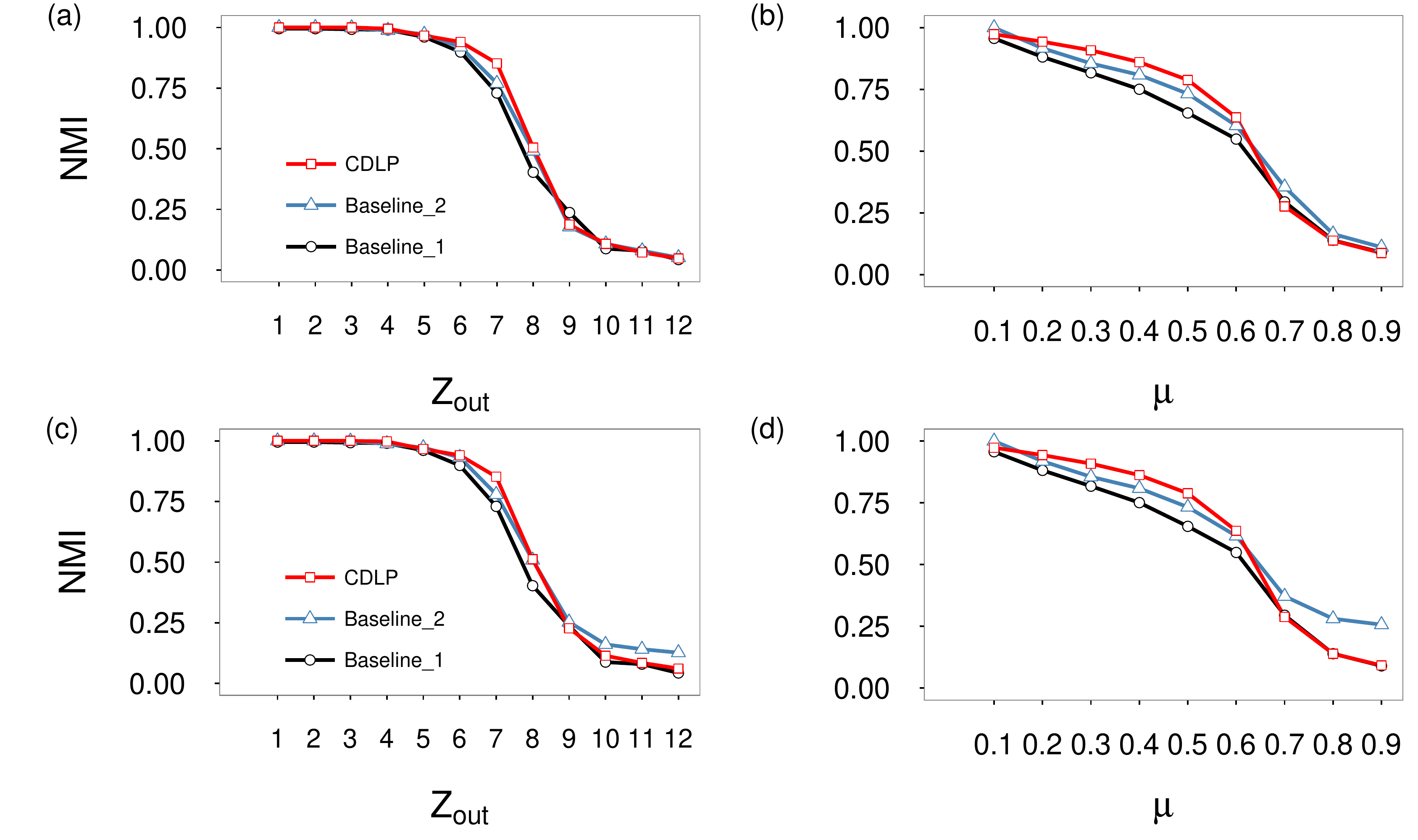}
\caption{The top row contains computed  averaged NMI value of baseline and proposed algorithms where modularity metric is applied in choosing step, for (a) GN networks with different $Z_{out}$ value and (b) LFR networks with different $\mu$ value. The bottom row is the replication of the top row,
except for the difference in the step of choosing optimal predicted network, where NMI metric is adopted, for (c) GN networks with different $Z_{out}$ value and (d) LFR networks with different $\mu$ value. The legend denotes the results on different objective algorithms, where black line and circle means the result of  baseline algorithm, red line and rectangle indicates the result of proposed algorithm, while the rest illustrates the second class of baseline.}\label{Fig:figure1}
\end{figure*}

For each artificial network with the same parameters, we generate 10 data sets, and all results are averaged on them.
As we can see from Fig. \ref{Fig:figure1}, although increasing $Z_{out}$ and $\mu$ yield different results, our proposed algorithm is the winner in most cases.
We can make the conclusions as follows: (1) Our method (the square shapes) and the second class of baseline (the triangle shapes) both outperform the first class of baseline (the circle shapes) in most cases, indicating that link prediction can be devoted to improved accuracy of community detection.
This experiment exhibits great significance of true network structure, and attaches great importance to inclusion of link prediction when detecting community structures.
(2) Since previous analysis shows that community detection with inclusion of link prediction will perform better, here we find out that different similarity measurement used in link prediction yields different results. As is shown in Fig. \ref{Fig:figure1}, result of  our proposed similarity measurement outperforms that of CN measurement under most circumstances. However, as community structures becomes fuzzy with increasing $Z_{out}$ and $\mu$ value, which makes it difficult to detect community partition, CN measurement is more suitable. This phenomenon is in accordance with the fact that fuzzy community structures undermine the efficiency of new measurement, as CN measurement is unaffected by community structures.
(3) Considering modularity metric can be computed in both artificial network and real world network, we apply the measure of modularity to choose the optimal combination of parameters for link prediction. Hence, we compare the result of applying modularity as choosing metric with that of applying NMI. From Fig. \ref{Fig:figure1}, it makes little difference about which metric to be applied for choosing model. In other words, in terms of choosing model,  two metrics show highly consistency in almost all networks. This also motivates us to think about the relationship between NMI and modularity, which has not been studied yet.

\section{Conclusions and Future work}\label{conclusion}
In this paper, we propose CDLP method to improve the existing community detection algorithms using link prediction from a new perspective. Compared with the two classes of baseline, we draw two main conclusions: (1) Link prediction can be devoted to detect more accurate community structures. (2) In link prediction step, our proposed novel indices outperform the CN measurement in most cases, illustrating that more accurate network structure leads to more accurate community structures. In addition, more credible community information can be devoted to the improvement of link prediction precision, so here comes a significant conclusion: link prediction and community structure are mutually beneficial, demonstrating a virtuous circle.
Based on our works, there are several attractive problems that are worthy of trying for future work, including generalization of the novel model to weighted and directed networks, and the relationship between modularity and NMI.

\bibliographystyle{cpb}
\bibliography{reference}

\end{document}